\documentclass{article}
\usepackage{PRIMEarxiv}
\usepackage[font=small, labelfont=bf]{caption}
\usepackage{float}
\usepackage{xcolor}
\usepackage{ragged2e}
\usepackage{ulem}
\usepackage{amsfonts}
\usepackage{amsmath}
\usepackage{fixmath}

\usepackage{color} 
\usepackage{subfigure}
\usepackage{graphicx}
\usepackage{dcolumn}
\usepackage{bm}

\newcommand{\beq}{\begin{eqnarray}}
\newcommand{\eeq}{\end{eqnarray}}
 \graphicspath{ {./FIG/} }
 \makeatletter
\def\blfootnote{\xdef\@thefnmark{}\@footnotetext}
\makeatother

\usepackage{todonotes}

\title{Fourier's law  breakdown for the  planar-rotor chain with long-range coupling}

\author{Henrique Santos Lima$^{1}$, Constantino Tsallis$^{2}$, Deniz Eroglu$^{3}$ and Ugur Tirnakli$^{4}$ }

\date{\today}
\begin{document}
\maketitle
\begin{abstract}
The thermal conductivity, $\kappa$, of a homogeneous chain of generically-ranged interacting planar rotors, more precisely the inertial $\alpha-XY$ model, is numerically studied with the coupling constant decaying as $r^{-\alpha}$. The conductance ($\sigma\equiv \kappa/L$) is calculated for typical values of  $\alpha$, temperature $T$ and lattice size $L$. For large $L$, the results can be  collapsed into an universal $q$-stretched exponential  form, namely    $L^{\delta_{\alpha}} \sigma \propto  e_{q_{\alpha}}^{-B_{\alpha}(L^{\gamma_{\alpha}}T)^{\eta_{\alpha}}}$, where $e_q^z\equiv [1+ (1-q)z]^{1/(1-q)}$. The parameters $(\gamma_{\alpha},\delta_{\alpha},B_{\alpha}, \eta_{\alpha})$ are $\alpha$-dependent, and $q_{\alpha}$ is the index of the $q$-stretched exponential. This form is achievable due to the ratio $\eta_{\alpha}/(q-1)$ being almost constant with respect to the lattice size $L$.
Two distinct regimes are identified: for $0 \le \alpha < 2$, the  Fourier law is violated, whereas for $\alpha\ge 2$, it is satisfied. These findings provide significant insights into heat conduction mechanisms in systems with long-range interactions.

\end{abstract}
\blfootnote{\textit{$^{1}$~Centro Brasileiro de Pesquisas Fisicas, Rua Xavier Sigaud 150, Rio de Janeiro-RJ 22290-180, Brazil.\\ 
E-mail: hslima94@cbpf.br }}

\blfootnote{\textit{$^{2}$~Centro Brasileiro de Pesquisas Fisicas and National Institute of Science and Technology of Complex Systems, Rua Xavier Sigaud 150, Rio de Janeiro-RJ 22290-180, Brazil \\
Santa Fe Institute, 1399 Hyde Park Road, Santa Fe, 
 New Mexico 87501, USA \\
Complexity Science Hub Vienna, Josefst\"adter Strasse 
 39, 1080 Vienna, Austria \\
E-mail: tsallis@cbpf.br}}

\blfootnote{\textit{$^{3}$~Faculty of Engineering and Natural Sciences, Kadir Has University, 34083, Istanbul, Turkey 
E-mail: deniz.eroglu@khas.edu.tr}}

\blfootnote{\textit{$^{4}$~Department of Physics, Faculty of Arts and Sciences, Izmir University of Economics, 35330, Izmir, Turkey 
E-mail: ugur.tirnakli@ieu.edu.tr}}

\maketitle
\section{Introduction}

The Fourier's law, which states that the heat flux within a material is proportional to the negative gradient of the temperature field, was proposed by J. Fourier in 1822~\cite{Fourier1822}. Today, it is intensively investigated as a prototype of nonequilibrium phenomena \cite{Lebowitz1978,Buttner1984,Maddox1984,Laurencot1998,Michel2003,Kawaguchi2005,Bernardin2005,Gruber2005,Bricmont20071,Bricmont20072,Wu2008,Gaspard2008, DubiVentra20091,DubiVentra20092}. For one-dimensional systems, the Fourier's law can be written as 
\begin{equation}
    J=-\kappa \frac{d\phi}{dx}\,,
    \label{hf}
\end{equation}
where $J$ is the heat flux, $d\phi/dx$ is the thermal gradient related with the temperature field $\phi=\phi(x,t)$ and $\kappa$ is the thermal {\it conductivity}, assumed to be finite. The temperature $T$ is defined as the spatial average of the temperature field in the stationary state. For instance, in a system with reservoirs $R_h$ and $R_l$ corresponding to the hot and cold reservoirs, respectively, the temperature is  $T=\frac{\phi_{st}(0)+\phi_{st}(L)}{2}$, where $\phi_{st}(x)$ is the temperature field at the stationary state. Here, $x=0$ corresponds to the point where the hot reservoir is located, while $x=L$ to the cold reservoir.


In principle, the thermal conductivity $\kappa$ is typically a function of intensive parameters such as temperature and pressure~\cite{Flumerfelt1969}. The law of linearity between heat flux and thermal gradient, as expressed by Fourier's law, is widely validated across various three-dimensional systems and even in some low-dimensional ones~\cite{Aoki2000, Dahr2011, LandiOliveira2014, TsallisLima2023, LimaTsallis2023,LimaTsallisNobre2023}.

However, lattice models can also depend on extensive parameters like the lattice size $L$. When this dependence becomes anomalous, the thermal conduction exhibits non-standard behavior, characterized by $\kappa\sim L^{\rho}$, where $\rho$ is the lattice size exponent. Anomalous $\kappa$ implies a breakdown of Fourier's law, as expressed by Eq.~\eqref{hf}, due to divergence or vanishing of thermal conductance ($\rho>0$ or $\rho<0$, respectively) in the thermodynamic limit. Interestingly, thermal conductivity can also be influenced by defects, as observed in materials such as carbon nanotubes~\cite{HanFina2011}. Anomalies in thermal conductivity can be seen in Refs.~\cite{Flumerfelt1969,YangZhang2010,Gers2010,LiuXu2012,Xu2016,Hurtado2016,Wu2017}. Anomalous transport phenomena in lattice systems with variable size $L$ provide valuable insights into these deviations. First-principle numerical simulations of linear chains of nearest-neighbor-coupled planar rotators have yielded significant findings regarding thermal conductivity, $\kappa$ and thermal conductance, $\sigma\equiv \kappa/L$, in spin models~\cite{HarrisGrant1988,Savin2005,Louis2006,Menaetal2020,Aziz2022,Kojimaetal2022,Chauhanetal2022}.

References~\cite{LiLiLi2015,LiLi2017} have shown that assuming the first and last rotators are in thermal contact with Langevin heat baths at slightly different temperatures $T_h$ and $T_l$,  $\sigma (T, L)$ follows a $q$-Gaussian form, a typical function of $q$-statistics~\cite{Tsallis1988,LimaTsallis2020,tsallisbook2023}, namely, $e_q^{-B\,(L^{1/3}T)^2}$ with $B>0$, $q \simeq 1.55$,  $e_q^z \equiv [1+(1-q)z]^{1/(1-q)}$, and $T \equiv \frac{T_h+T_l}{2}$. For large values of  $L^{1/3}T$, the thermal conductance $\sigma$ behaves according to a power law,  $\sigma\sim (L^{1/3}T)^{-\frac{2}{q-1}}$, which definitively violating Fourier's law. This is due to $\kappa$ scaling as $\kappa\propto L^{1-\frac{2}{3}\frac{1}{q-1}}\sim L^{-0.21}$. At higher temperature ranges, a more general expression known as the $q$-stretched exponential form is considered. The thermal conductances of the one-dimensional $n$-vector models~\cite{Stanley1968} are well fitted by the functional form:
 \begin{equation}
     \sigma(T;L)=A\,e_q^{-B(L^{\gamma}T)^{\eta}}
 \end{equation}
 which has been verified that for $n=1,2,3$ these models exhibit normal heat conduction~\cite{TsallisLima2023,LimaTsallis2023,LimaTsallisNobre2023}. 

 Due to their complexity, generic-range spin models are rarely investigated in the literature regarding nonequilibrium properties~\cite{OlivaresAnteneodo2016}. The main characteristic of these models is the potential that decays as a power law with the distance between constituents. These models are known for exhibiting abnormalities such as the inequivalence between microcanonical and canonical ensembles in terms of equilibrium properties, in the limit of very long-range interactions. For example, gravitational systems, which are common long-ranged systems,  present negative specific heat and extensivity breaking with the framework of standard statistical mechanics~\cite{Julien2001, Pik2014}.

Since the standard $XY-$model is naturally short-ranged, it is possible to generalize this model to cover generic-range interactions, assuming the exchange coupling matrix  $j_{ij}$ is proportional to $1/r_{ij}^{\alpha}$ ($\alpha \ge 0$), where $r_{ij}\equiv |i-j|$ is the distance between rotators and $\alpha$ is the exponent that dictates the type of interaction~\cite{JundTsallis1995,AnteneodoTsallis1998}. This model is called the $\alpha-XY$ model. In the  $\alpha\to \infty$ limit, it recovers the above-mentioned nearest-neighborhood case, while in the $\alpha=0$ limit, it recovers the mean field  model. The influence of arbitrary $\alpha$ in this peculiar heat transport phenomenon was explored in \cite{OlivaresAnteneodo2016}, where a special value of $\alpha$ was detected. Larger than this value, the Fourier's law is satisfied at high temperatures, whereas it is violated in all cases smaller than the critical value.
 This article numerically investigates the thermal conductivity of the classical inertial $\alpha-XY$ for various values of $\alpha$, as well as determining where Fourier's law holds and where it is broken. 
For clarity, let us anticipate at this stage our present main conclusion, namely that the results can, in all cases, be collapsed in the following universal form:  $L^{\delta_{\alpha}}\sigma(T,L; \alpha)= A_{\alpha}e_{q_{\alpha}}^{-B_{\alpha}(L^{\gamma_{\alpha}}T)^{\eta_{\alpha}}}$, where $(\gamma_{\alpha},\delta_{\alpha}, A_{\alpha}, B_{\alpha})$ are $\alpha$-dependent  non-negative coefficients and $q_{\alpha}$ is the index of the $q$-stretched exponential. We also investigate the threshold where, from that, the Fourier's law holds.
 
\section{Model}
The Hamiltonian of the classical inertial one-dimensional $\alpha-XY$ is given by
\begin{equation}
\label{ham}
    \mathcal{H}=\sum_{i=1}^L\frac{p_i^2}{2}+\frac{\varepsilon}{2\tilde L}\mathop{\sum_{j=1}^{L}\sum_{i=1}^{L}}_{j\neq i}\frac{1-\cos(\theta_i-\theta_j)}{r_{ij}^{\alpha}}\,,
\end{equation}
where $p_i$ and $\theta_i$ are the angular momenta and coordinates, respectively, and $\varepsilon$ is the exchange coupling. The moment of inertia of the rotators is set to unity without loss of generality. The factor $\tilde{L}$,
\begin{equation}
    \tilde L\equiv\frac{1}{L} \mathop{\sum_{i,j}}_{j\neq i}r_{ij}^{-\alpha}\,,
\end{equation}
  is introduced to make the Hamiltonian extensive  \cite{JundTsallis1995}. Notice that, for $\alpha=0$, $\tilde L=L-1\sim L$, and for $\alpha\to \infty$, 
  $\tilde L =2$. For the $d$-dimensional case, we change $\tilde{L} \to \tilde{N}$,  where $N$ is no longer the linear lattice size. In this notation, for $\alpha=0$,  $\tilde{N}=N-1\sim N$, and for $\alpha\to \infty$,  $\tilde{N} =2d$, where $d$ is the dimension of the system. The general expression for $\tilde{N}$ is given by a relation proportional to
\begin{equation}
    \ln_{\alpha/d}{N}\equiv \frac{N^{1-\alpha/d}-1}{1-\alpha/d}\,,
    \label{log}
\end{equation}
which  behaves as $\tilde{N}\sim N^{1-\alpha/d}$ for $0\le\alpha/d<1$, presents a logarithmic divergence as $\tilde{N}\sim \ln{N}$ for $\alpha/d=1$. For $\alpha/d>1$, $\tilde{N}$ approaches a finite value in the thermodynamic limit. Thus, a critical value $\alpha_c/d=1$ is identified; below this value, the system exhibits very long-range interactions, and above it, the system transitions to other types of interactions. Let us clarify that for $0\le \alpha/d<1$, the system is very long-ranged, while for $1<\alpha/d<\infty$ the system is long-ranged. The particular case of nearest-neighbors is the only one that is consistent with the definition of short-range interaction. The main explanation is because only for $\alpha/d\to \infty$ we can guarantee that all momenta of its distribution are finite, and beyond that, for all cases in which  $\alpha/d<\infty$, the power law behavior of the interactions can not be neglected.

In this article, the one-dimensional system is considered, and hence, the lattice size will be denoted as $L$ from now on. Therefore, $ \tilde{L}$ represents the adjusted notation. The equations of motion for the Hamiltonian described by  Eq. \eqref{ham}, with the addition of a Langevin heat bath where only the first and the last particles are coupled to the heat bath at  temperatures $T_h$ and $T_l$ ($T_h>T_l$), respectively, are given by
\begin{align}
\begin{split}
\dot \theta_i&=p_i\,\,\,\text{for $i=1, \dots, L$}\\
    \dot p_1&=-\gamma_h p_1+F_1+\sqrt{2\gamma_h T_h}\eta_h(t)\\
    \dot p_i &=F_i \,\,\,\text{for $i=2,\dots, L-1$}\\
    \dot p_L&=-\gamma_l p_L+F_L+\sqrt{2\gamma_l T_l}\,\eta_l(t)\\
    \end{split}
\end{align}
where the Boltzmann constant is set to unity. The generalized force (torque) components  are
\begin{equation}
    F_i\equiv -\frac{\varepsilon}{\tilde L}\sum_{j\neq i} \frac{\sin(\theta_i-\theta_j)}{r_{ij}^{\alpha}}
\end{equation}
and $\eta_{h}$ and $\eta_{l}$ are Gaussian white noises with the following correlations,

\begin{align}
\begin{split}
\langle \eta_{h}(t)\eta_{h}(t')\rangle&=\langle \eta_{l}(t)\eta_{l}(t')\rangle=\delta(t-t')\\ 
\langle \eta_{h}(t)\eta_{l}(t')\rangle&=0
    \end{split}
\end{align}

To measure the thermal conductance, the macroscopic  heat flux along the lattice, denoted as $\mathcal{J}$, is defined via  the continuity equation for each particle $\frac{d }{dt}\mathcal{H}_i=-\sum_{j\neq i}J_{ij}$ as in \cite{OlivaresAnteneodo2016}, such that
\begin{equation}
    J_{ij}\equiv \frac{\varepsilon}{2\tilde L}(p_i+p_j)\frac{\sin(\theta_i-\theta_j)}{r_{ij}^{\alpha}}.
\end{equation}
The average right-hand-side flux of particle-$i$, denoted $\mathcal{J}_i$, is defined as follows:
\begin{equation}
    \mathcal{J}_i\equiv \left \langle \sum_{j>i}J_{ij}\right\rangle\,.
\end{equation}
Note that the absolute value of the average left-hand-side flux equals the right-hand-side flux in the stationary state. Therefore, defining only one flux is sufficient for simplicity. Finally, the heat flux $\mathcal{J}$ is expressed as the average over the bulk particles, namely $\mathcal{J}\equiv \langle \mathcal{J}_i \rangle_{\text{bulk}}$, where the bulk is defined as all particles which are not in contact with the thermal baths. Thus,  the thermal conductance can be defined  as:
\begin{equation}
    \sigma=\frac{\mathcal{J}}{T_h-T_l}=\frac{\mathcal{J}}{2\Delta T}\equiv \frac{\kappa}{L}.
    \label{cond}
    \end{equation}
 To solve the equations of motion described by Eq.~\eqref{ham}, we utilize the Velocity Verlet algorithm \cite{Verlet1967} with a time step size of $dt = 0.01$, damping coefficients $\gamma_h = \gamma_l = 1.0$, interaction strength $\varepsilon = 2$, and temperatures $T_{h/l} = T(1 \pm \Delta)$, where $\Delta \equiv 0.125$, ensuring $T = (T_h + T_l)/2$.
 Initially, the coordinates and momenta are set to zero. The simulations run  from $t=0$ to the maximum time $t_{max}=1.65 \times 10^9$ ($1.65\times 10^{10}$ in the $L=20$ cases). The heat flux is averaged over 40 experiments for $L=20$ and $20$ experiments for $L=35,50$, each consisting of $4\times 10^8$ time steps. 
\section{Results}

\begin{figure*}[htb]
	\centering
	\includegraphics[width=17cm]{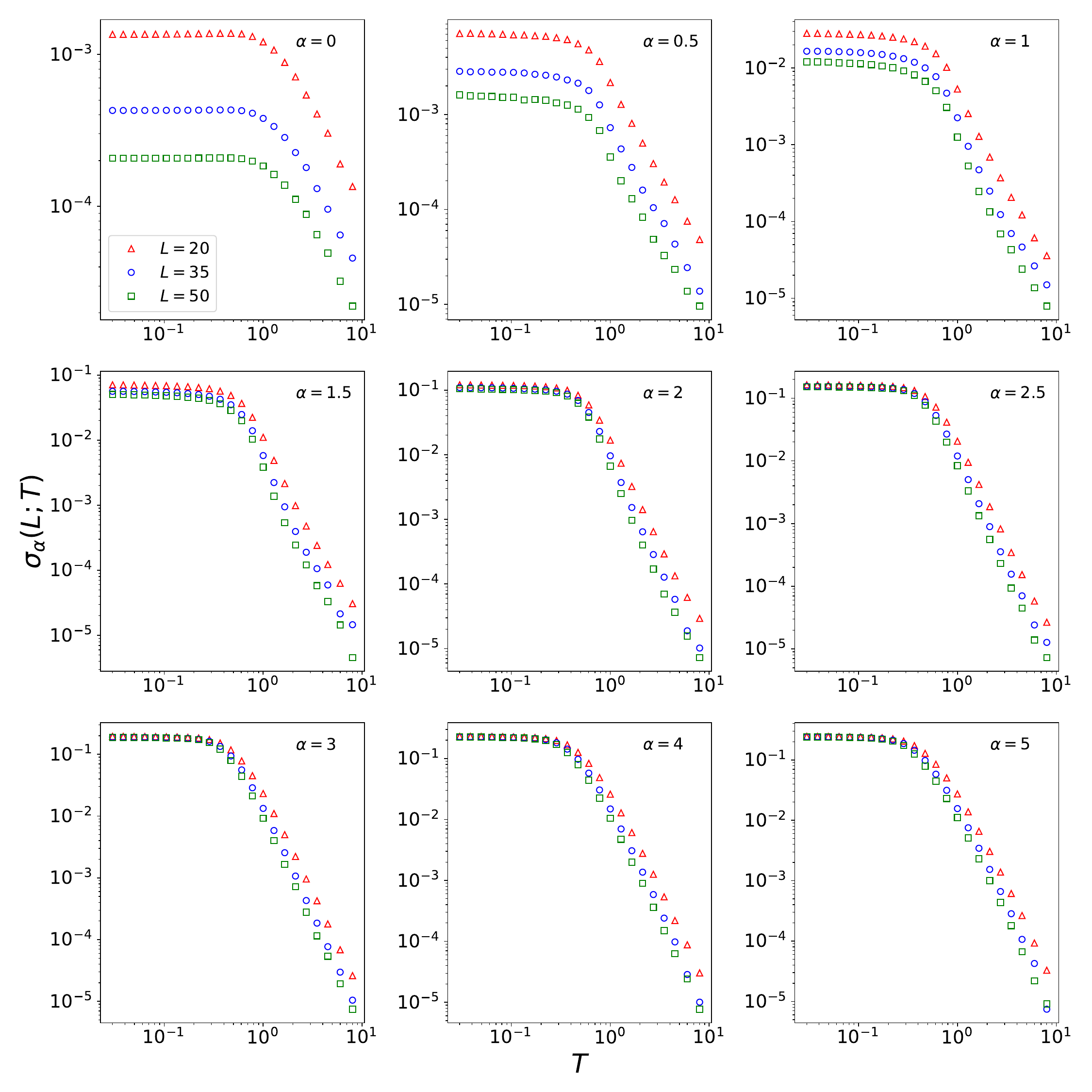}
	\caption{The plot of thermal conductance versus $T$ for $\alpha=0,0.5,1,1.5,2,3,4,5$ in log-log scale, for $L=20$ (red), $L=35$ (blue), and $L=50$ (green).  }
	\label{conductance}
\end{figure*}

\begin{figure*}[htb]
	\centering
	\includegraphics[width=17cm]{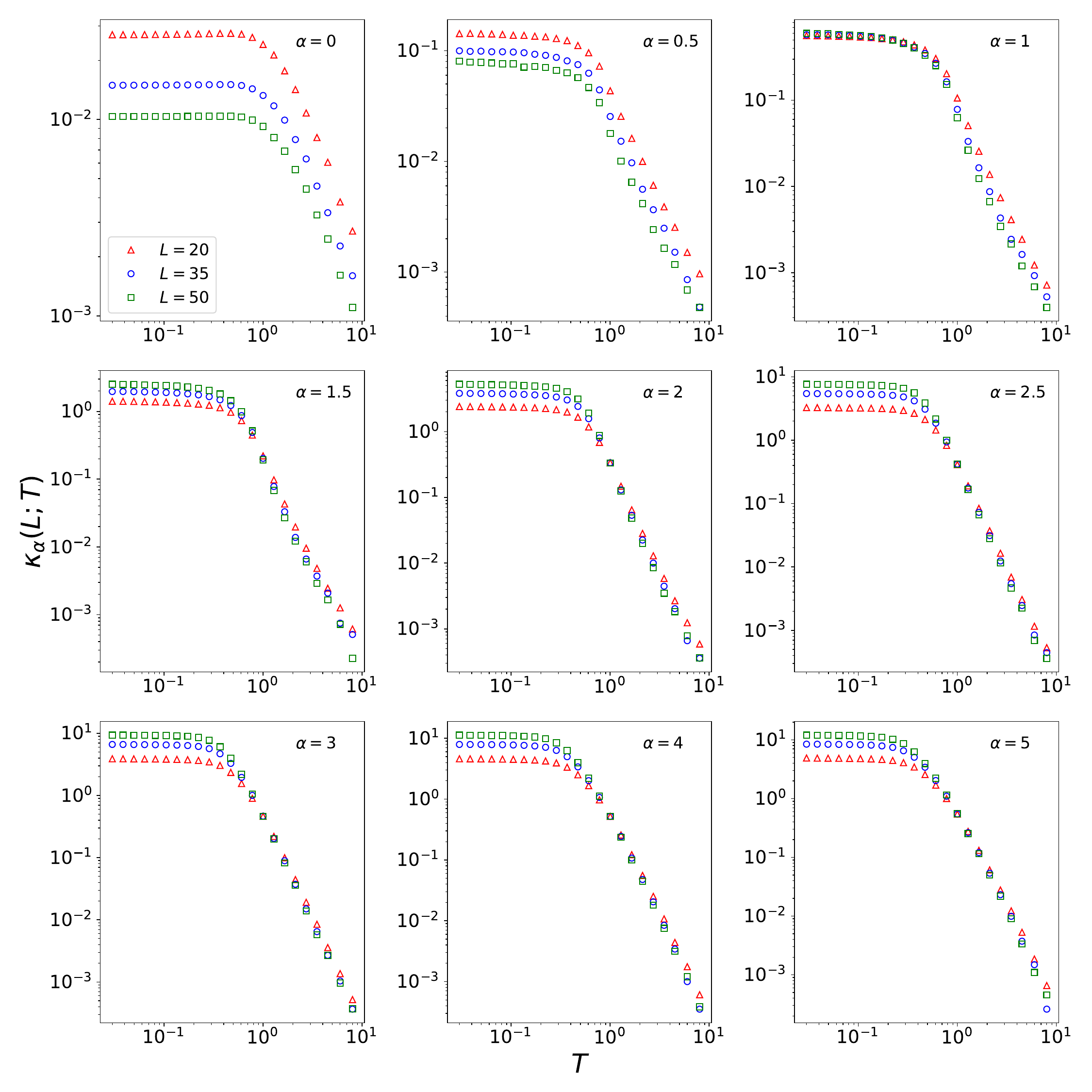}
	\caption{The plot of thermal conductivity versus $T$ for $\alpha=0,0.5,1,1.5,2,3,4,5$ in log-log scale, for $L=20$ (red), $L=35$ (blue), and $L=50$ (green). }
	\label{conductivity}
\end{figure*}

Fig.~\ref{conductance} illustrates that for $\alpha=0$, the thermal conductance varies substantially with the lattice size, and it decreases as $\alpha$ increases (notice that we use a different scale for $\alpha=0$). For $\alpha\ge 2$, the low-temperature regime of the thermal conductance appears to be independent of the lattice size, indicating that the thermal conductivity is ballistic in this regime. The threshold $\alpha=1$ seems to be almost uniformly dependent on the lattice size for all regimes. However, the logarithmic divergence in this case makes the high-temperature regime to exhibit an inflection close to $T=1$. The Fig.~\ref{conductivity} highlight this phenomenon. The divergence effect also disturbs their neighbors as we notice in the cases $\alpha=0.5$ and $\alpha=1.5$. Due to this dependence on the lattice size in the low temperature regime, it is reasonable  to assume the scaling of $\sigma_{\alpha} \sim L^{\delta_{\alpha}}$, where $\delta_{\alpha}$ is an exponent that depends on $\alpha$. This exponent decreases as $\alpha$ increases and it appears to vanish for $\alpha >2$.
 
Also, in Fig.~\ref{conductivity}, for $0\le\alpha<2$, the high-temperature regime is dependent on the lattice size, while out of this interval, the converse is noticed. Mainly for $\alpha=2$, the thermal conductivity substantially diminishes this dependence, as well as the inflection point almost disappears. The case $\alpha\to\infty$ was previously studied in Ref.~\cite{TsallisLima2023} and it exhibits similar behavior as in $\alpha\ge 2$ cases. This dependence justifies the scaling of the temperature with $L^{\gamma_{\alpha}}$, where $\gamma_{\alpha}$ also depends on $\alpha$.

The general expression of the thermal conductance is finally obtained as follows:
\begin{equation}
\label{sigma}
    \sigma_{\alpha}(T;L)=L^{-\delta_{\alpha}}A_{\alpha}e_{q_{\alpha}}^{-B_{\alpha}(L^{\gamma_{\alpha}}T)^{\eta_{\alpha}}} \,,
\end{equation}
where $A_{\alpha}$, $B_{\alpha}$, $\delta_{\alpha}$, and $\gamma_{\alpha}$ are non-negative parameters, $\eta_\alpha >2$, and $q_\alpha>1$. In the $L^{\gamma_\alpha}T \to \infty$ limit, this expression leads to
\begin{eqnarray}
\sigma_{\alpha}(T;L) &\sim& A_\alpha
L^{-\delta_\alpha}[B_{\alpha}(L^{\gamma_{\alpha}}T)^{\eta_{\alpha}}]^{1/(1-q_\alpha)}  \nonumber \\
&\propto& L^{-\bigl[\delta_\alpha + \frac{\gamma_\alpha \eta_\alpha}{q_\alpha -1}\bigr]}\,T^{-\frac{\eta_\alpha}{q_\alpha -1}}.
\label{scal}
\end{eqnarray}

\begin{figure*}[htb]
  \includegraphics[width=5.8cm]{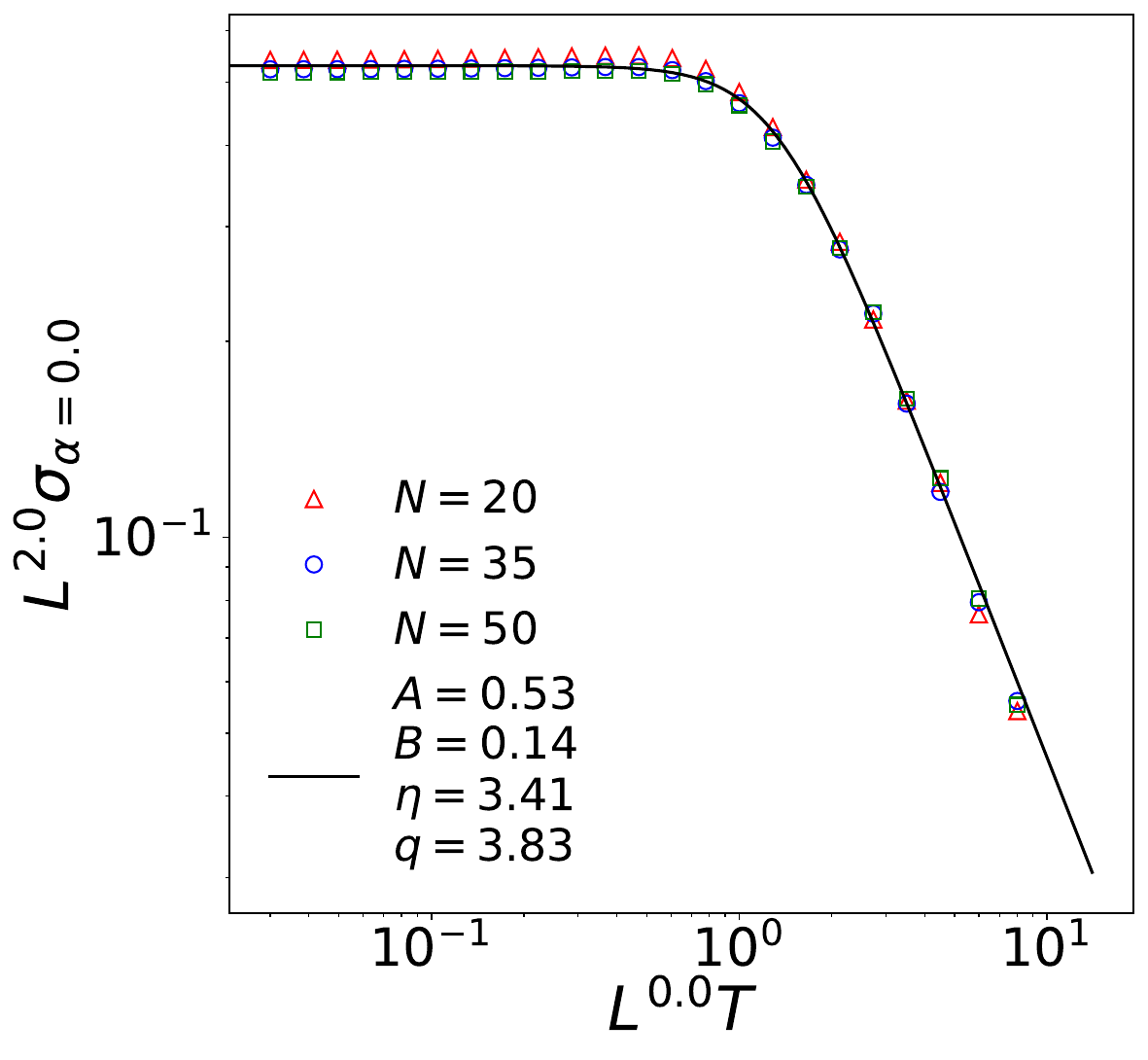}
 \includegraphics[width=5.8cm]{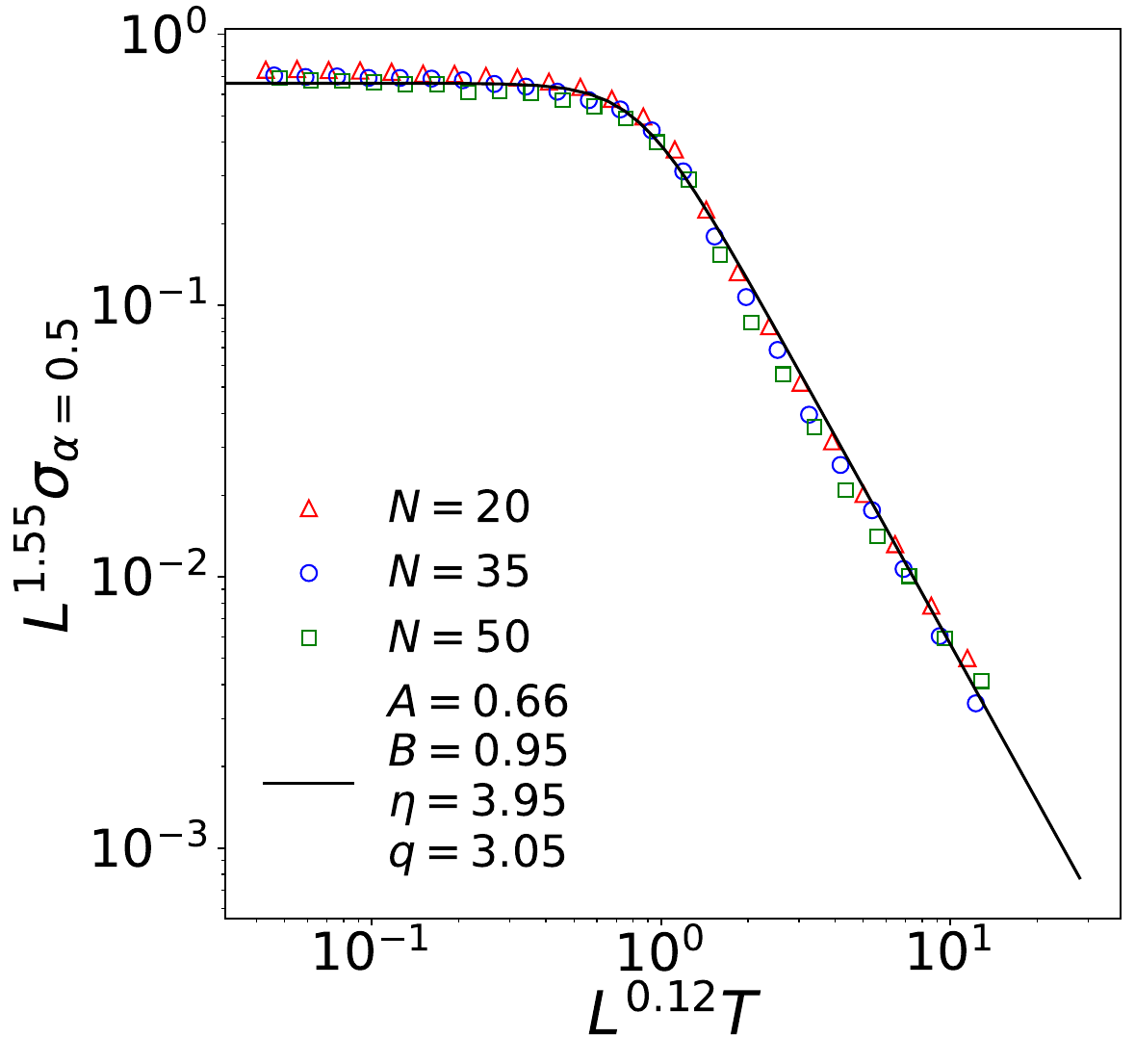}
 \includegraphics[width=5.8cm]{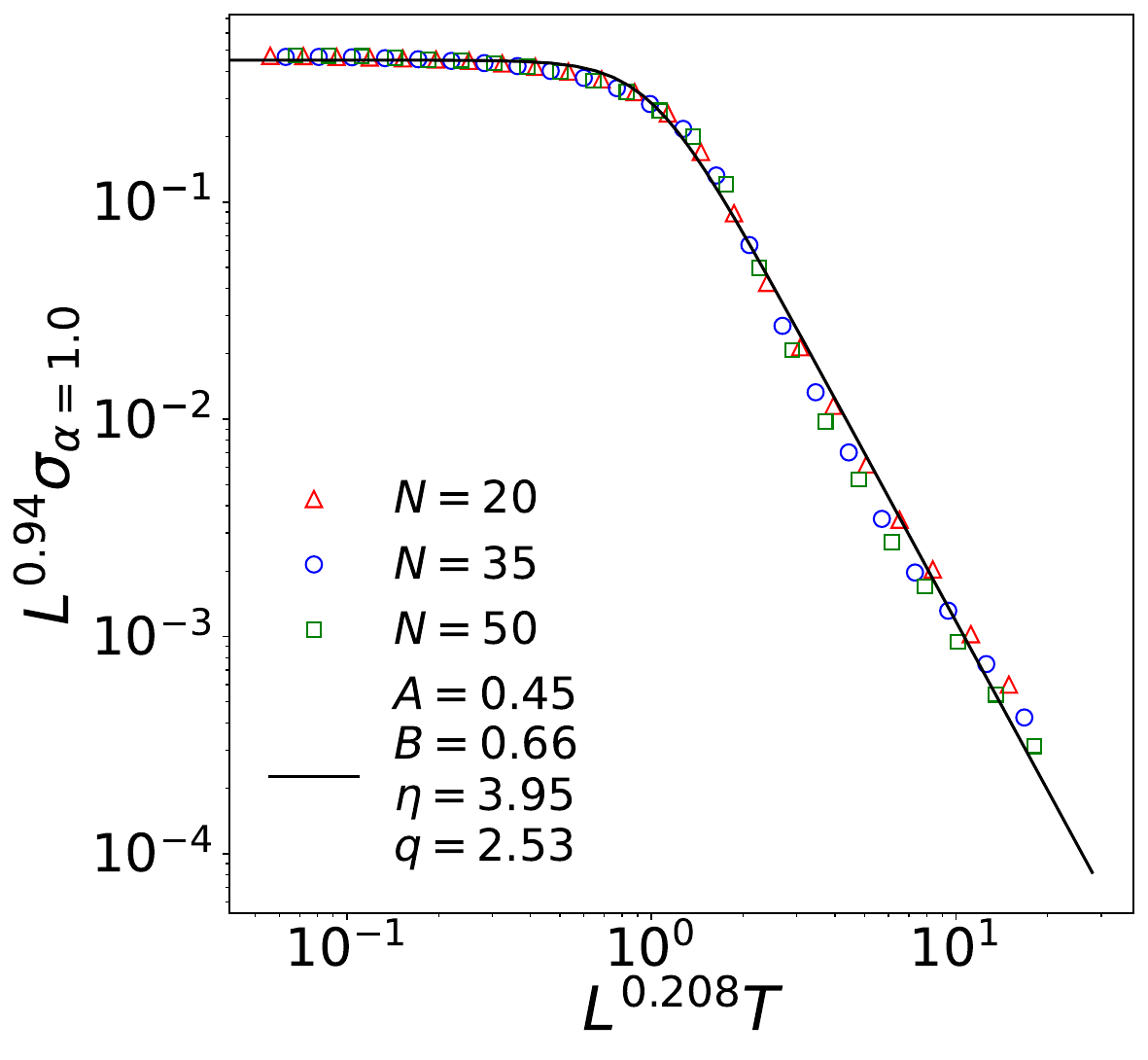}
 \includegraphics[width=5.8cm]{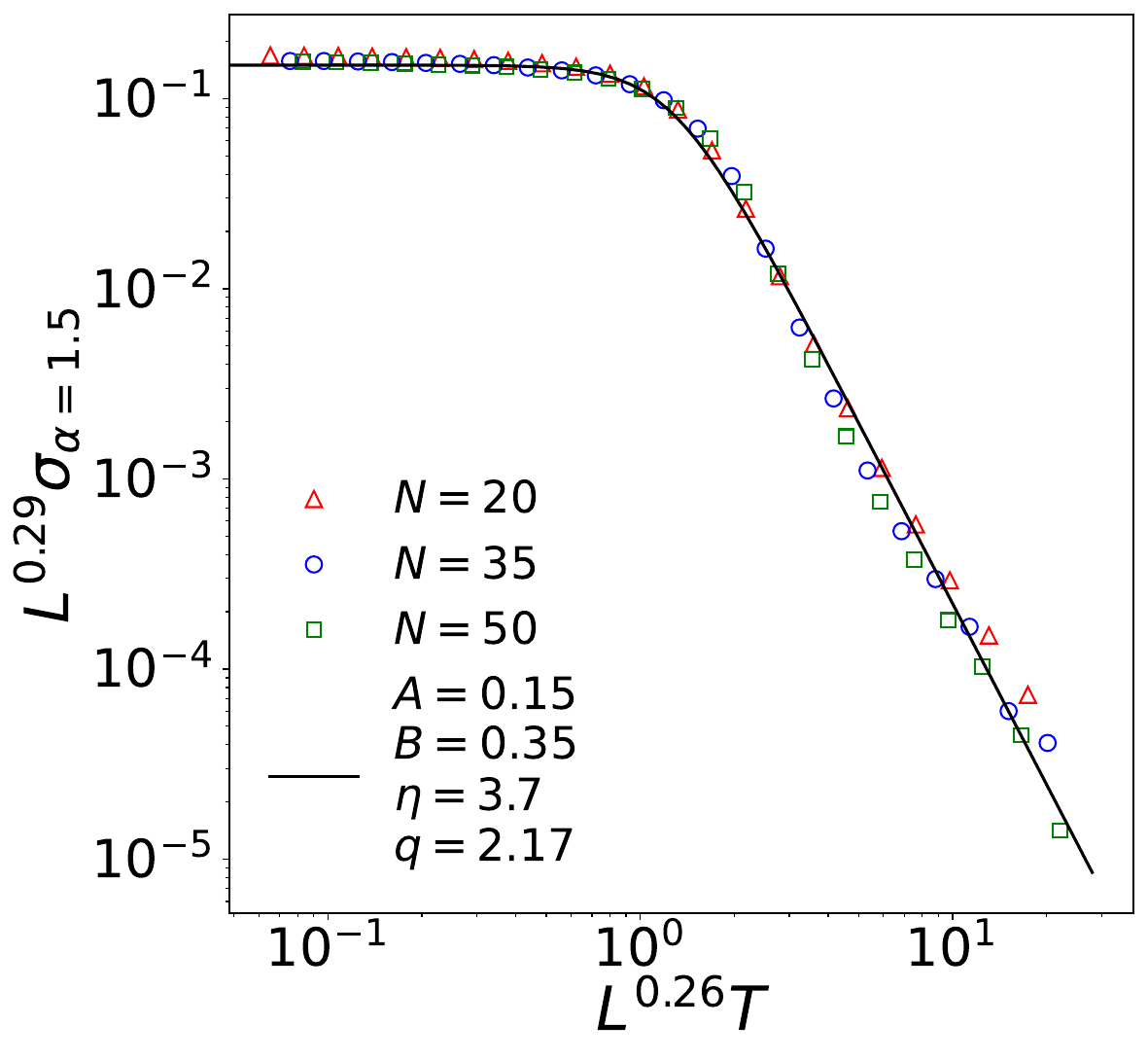}
 \includegraphics[width=5.8cm]{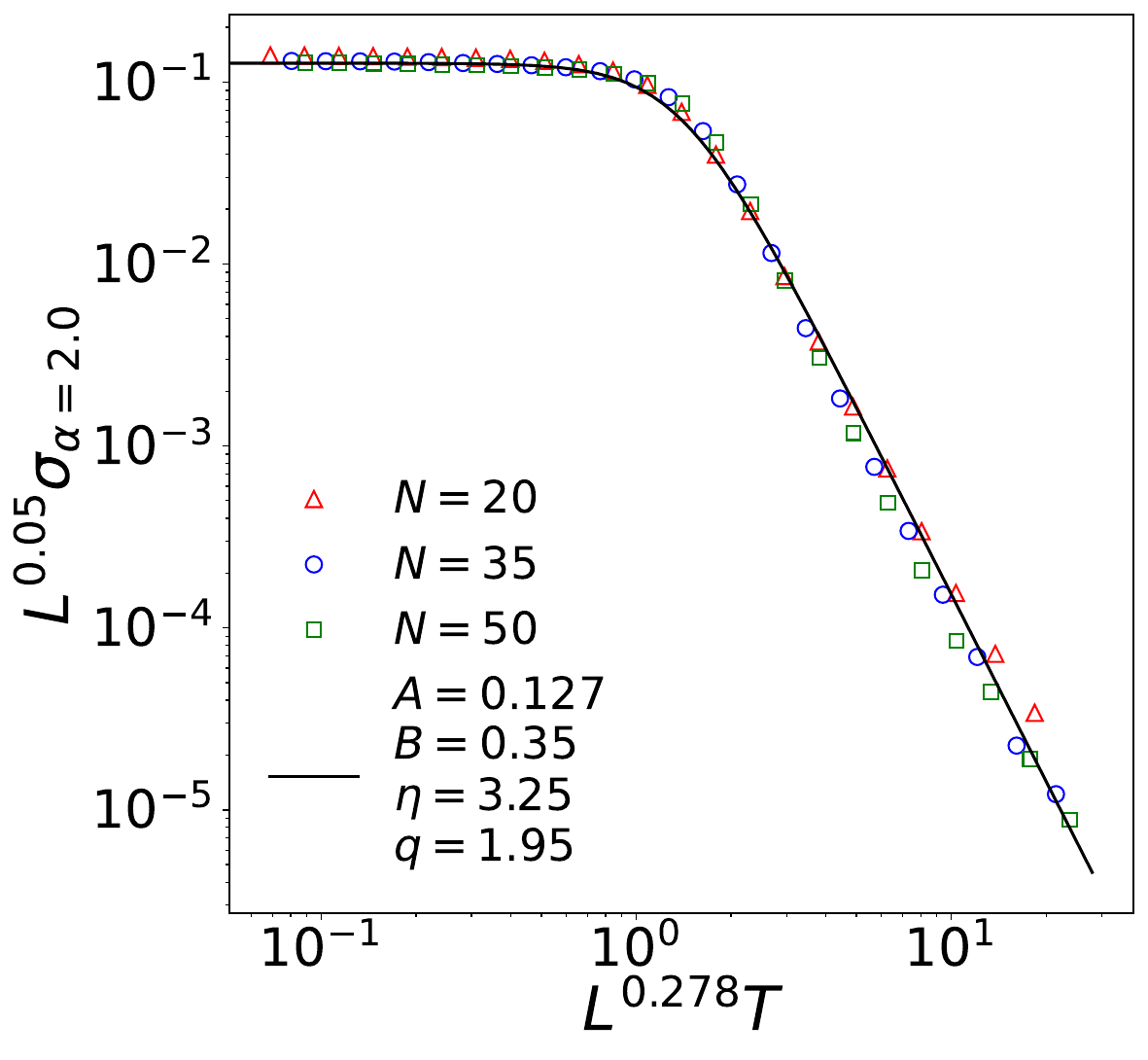}
 \includegraphics[width=5.8cm]{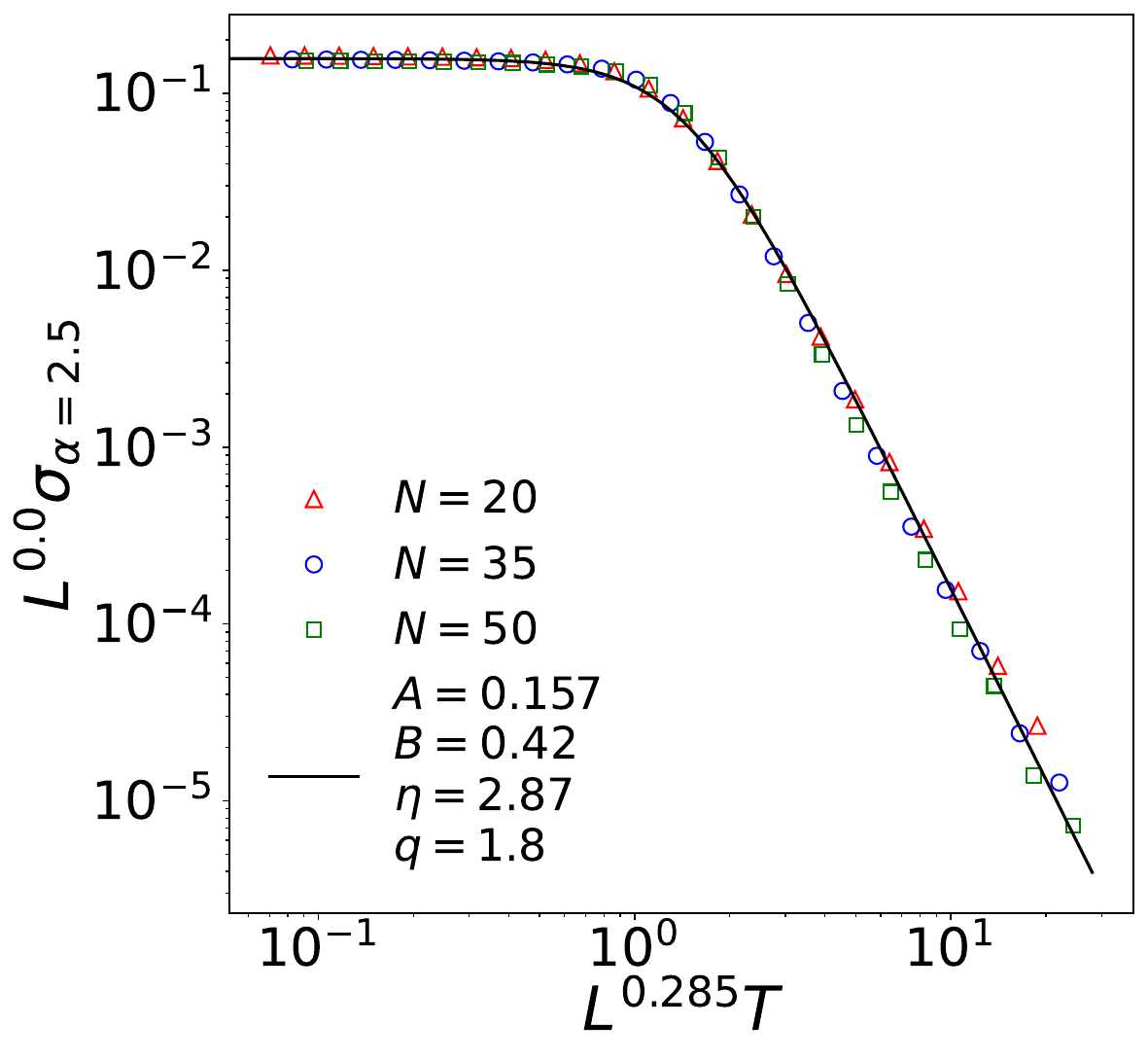}
 \includegraphics[width=5.8cm]{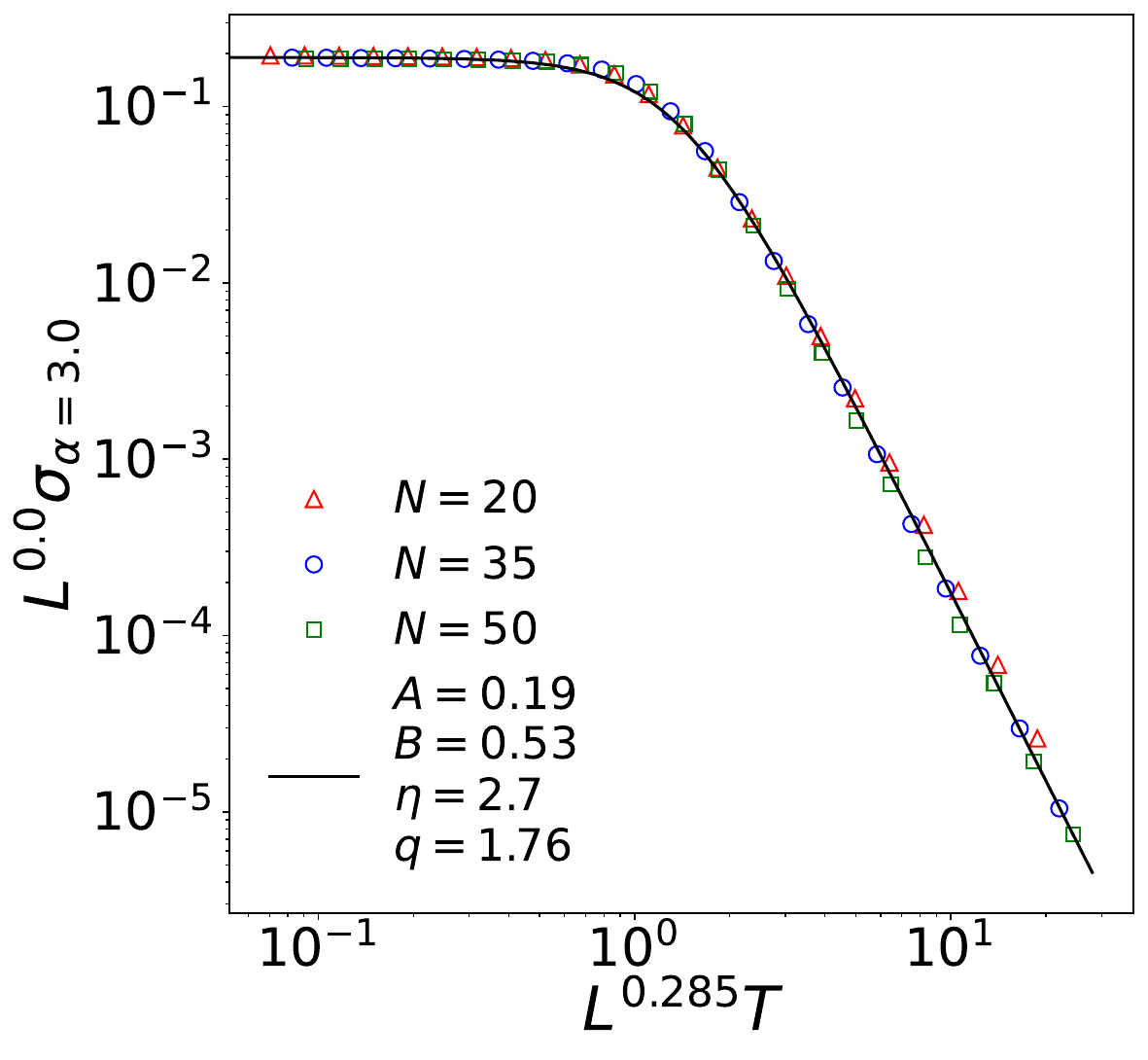}
	\includegraphics[width=5.8cm]{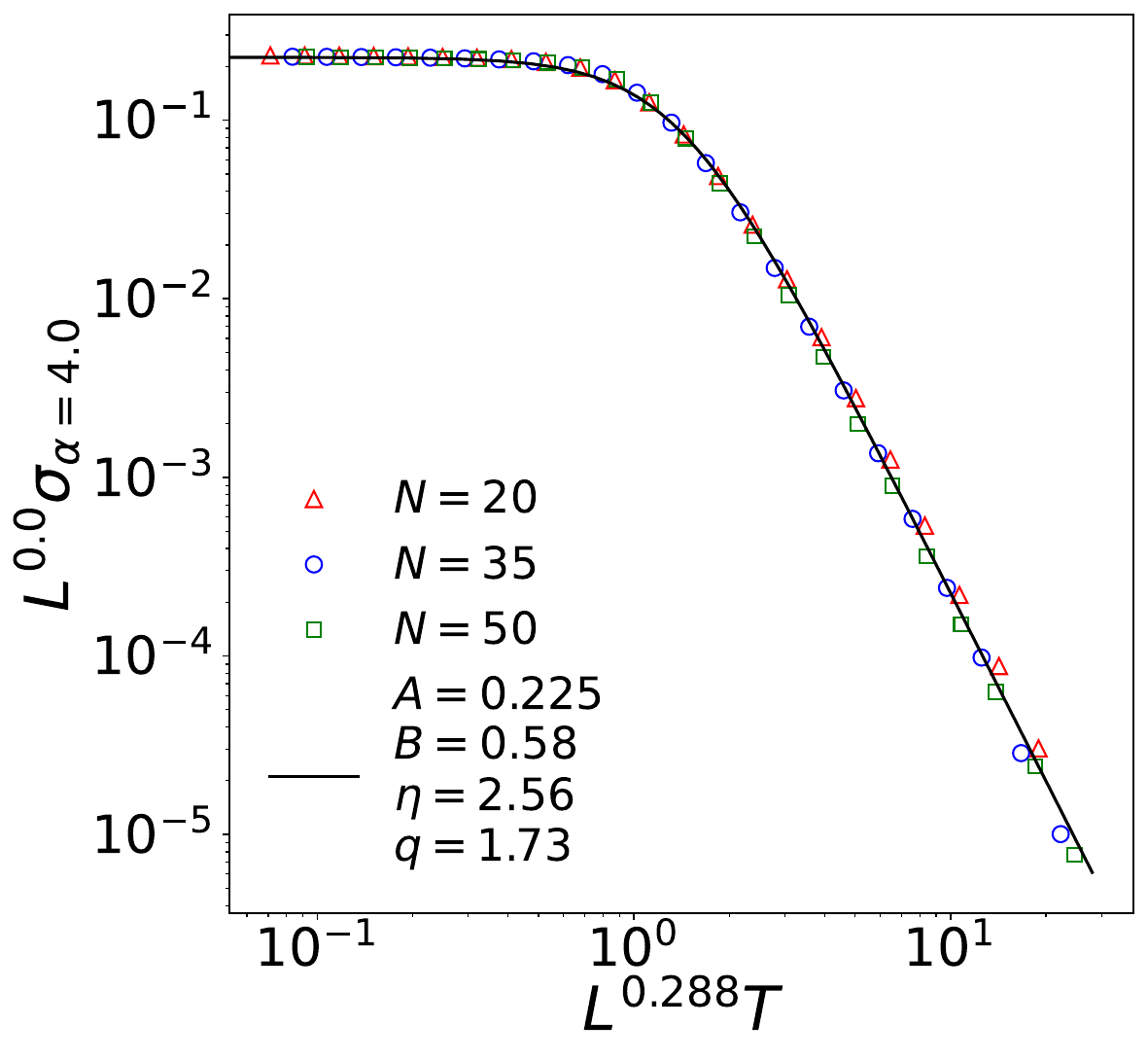}
    \includegraphics[width=5.8cm]{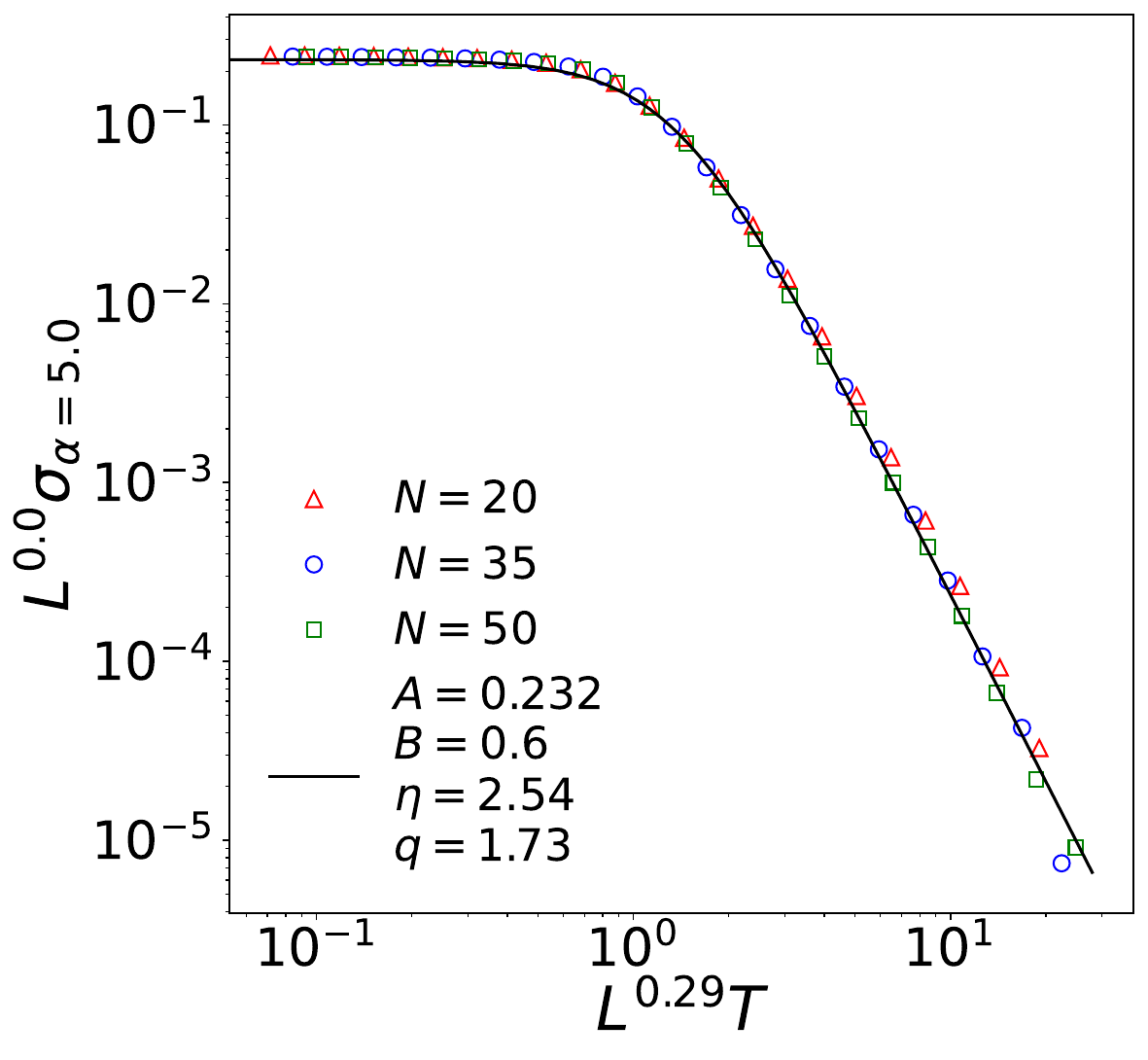}    
		\caption{ The plot of scaled thermal conductance versus $L^{\gamma}T$ for $\alpha=0,0.5,1,1.5,2,3,4,5$ in log-log scale, for $L=20$ (red), $L=35$ (blue), and $L=50$ (green). The black solid curve is a function as in Eq. ~\eqref{sigma}.}
	\label{collapse}
\end{figure*}

In Fig.~\ref{collapse}, we notice that the scaled thermal conductance, $L^{\delta_{\alpha}}\sigma_{\alpha}$ appears to behave as constant at the low-temperature regime, therefore, at this limit, $\sigma_{\alpha}\sim L^{-\delta_{\alpha}}$. As $\alpha$ increases, the $\delta_{\alpha}$ values decrease rapidly, and for $\alpha\ge2$ it almost vanishes. For $\alpha>2$, this parameter starts to be zero. Particularly, for the mean-field case ($\alpha=0$), the thermal conductivity decays as $\kappa_{\alpha=0}\propto L^{-1}$, which is invariant with respect to the Kac factor, in agreement  with the results in Ref~\cite{Defaveri2022}.

%

For the high-temperature regime at the thermodynamic limit, the thermal conductance starts to exhibit a different scaling, as in Eq.~\eqref{scal}. The particular cases $\alpha=2,2.5,3,4,5$ exhibit a scaling $\delta_\alpha+\gamma_\alpha \eta_\alpha/(q_\alpha-1)\sim 1$, in other words, $\sigma_{\alpha}\equiv \kappa_{\alpha}/L\sim L^{-1}$, indicating that $\kappa_{\alpha}\sim L^{0}$. The large-temperature asymptotic exponent of $\sigma_{\alpha}$ is the same of $\kappa_{\alpha}$, which yields $\sigma_{\alpha}\sim L^{-\eta_{\alpha}/(q_{\alpha}-1)}\sim \kappa_{\alpha}$ .

\begin{figure*}
	\centering
 \includegraphics[width=5cm]{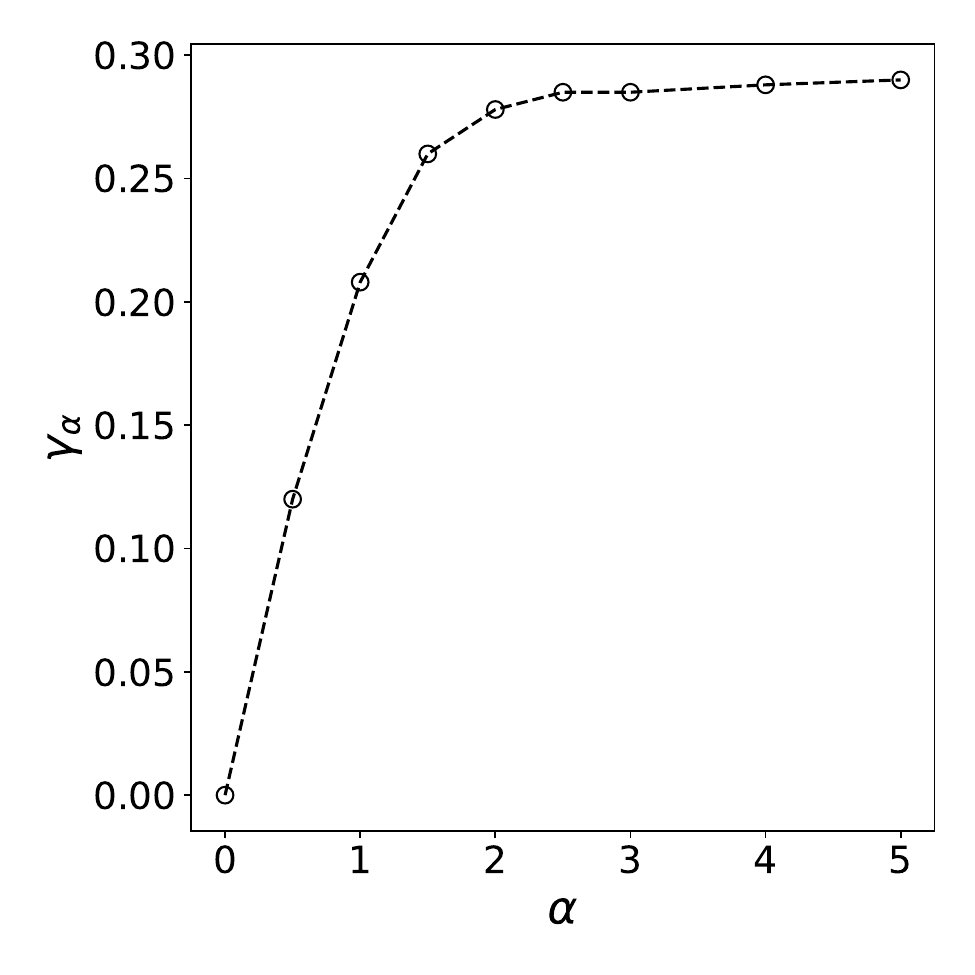}
 \includegraphics[width=5cm]{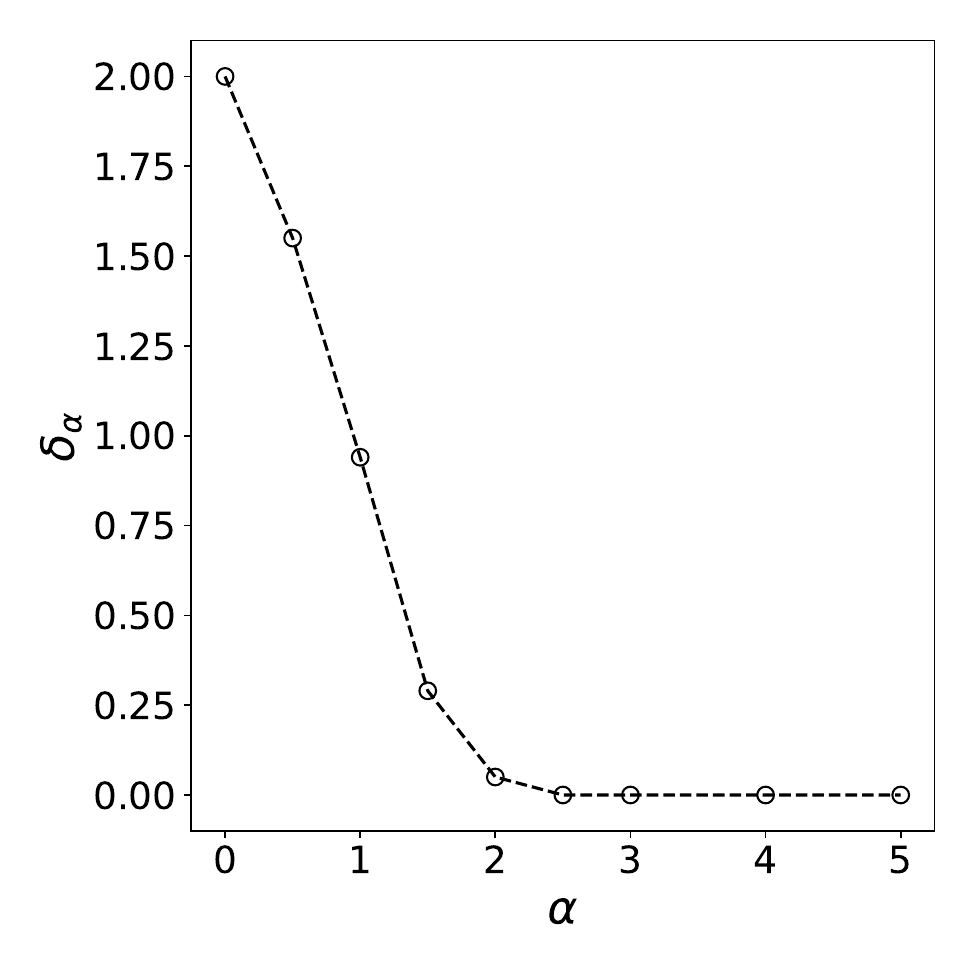}\\
 \includegraphics[width=5cm]{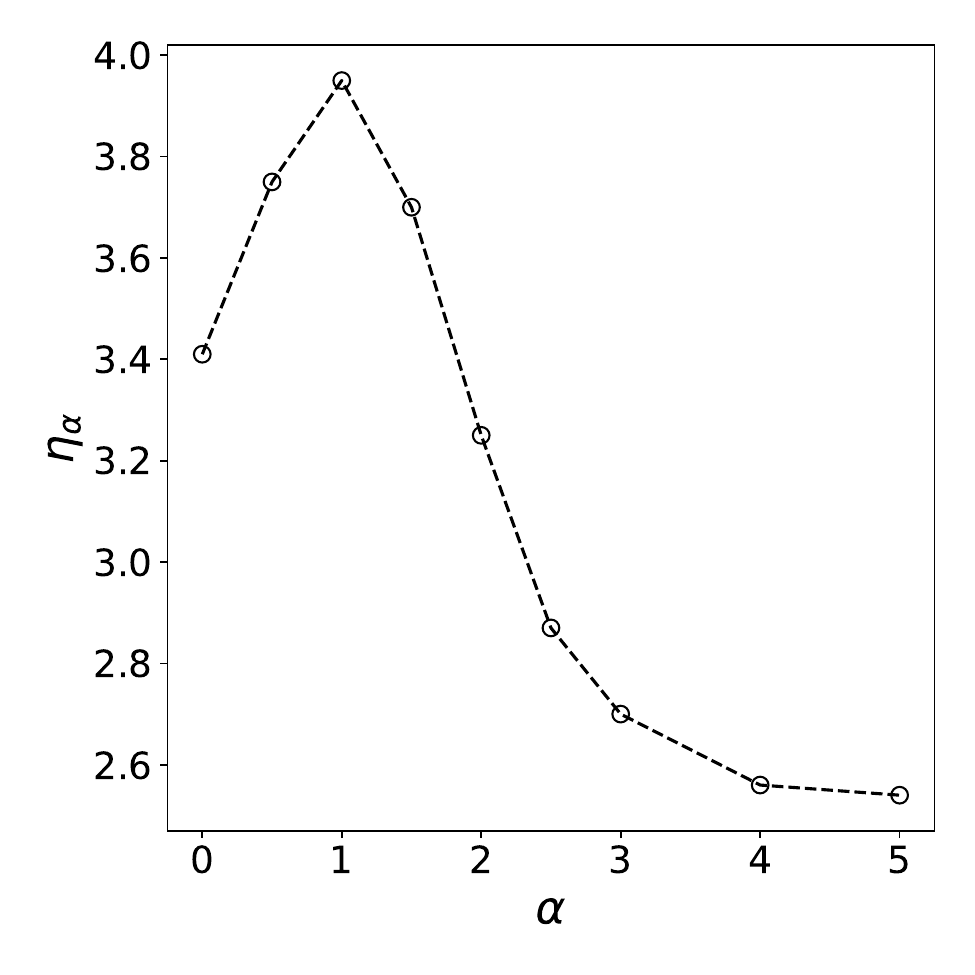}
 \includegraphics[width=5cm]{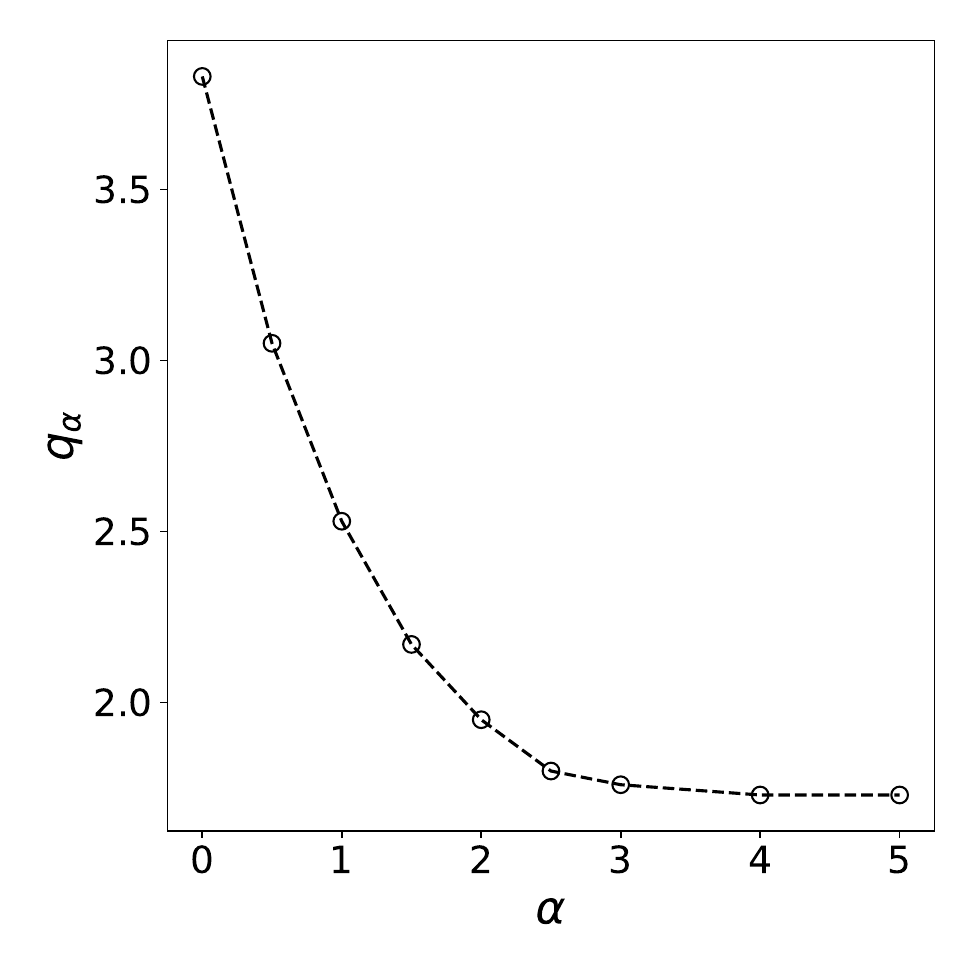} 
	\caption{
 Plot of  $(\gamma,\delta,\eta,q)$ as functions of $\alpha$. The polygonal solid lines are  guides to the eye.
 }
	\label{alpha}
\end{figure*}

\begin{figure*}
	\centering
 \includegraphics[width=5cm]{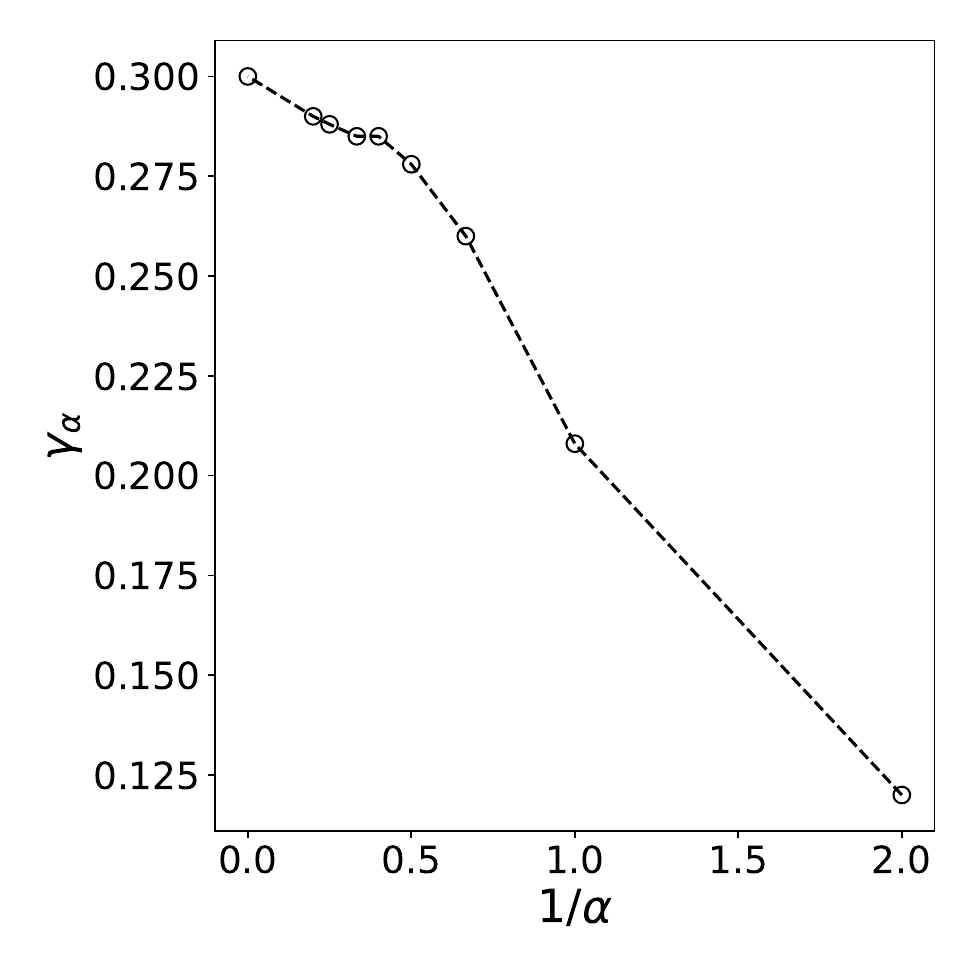}
 \includegraphics[width=5cm]{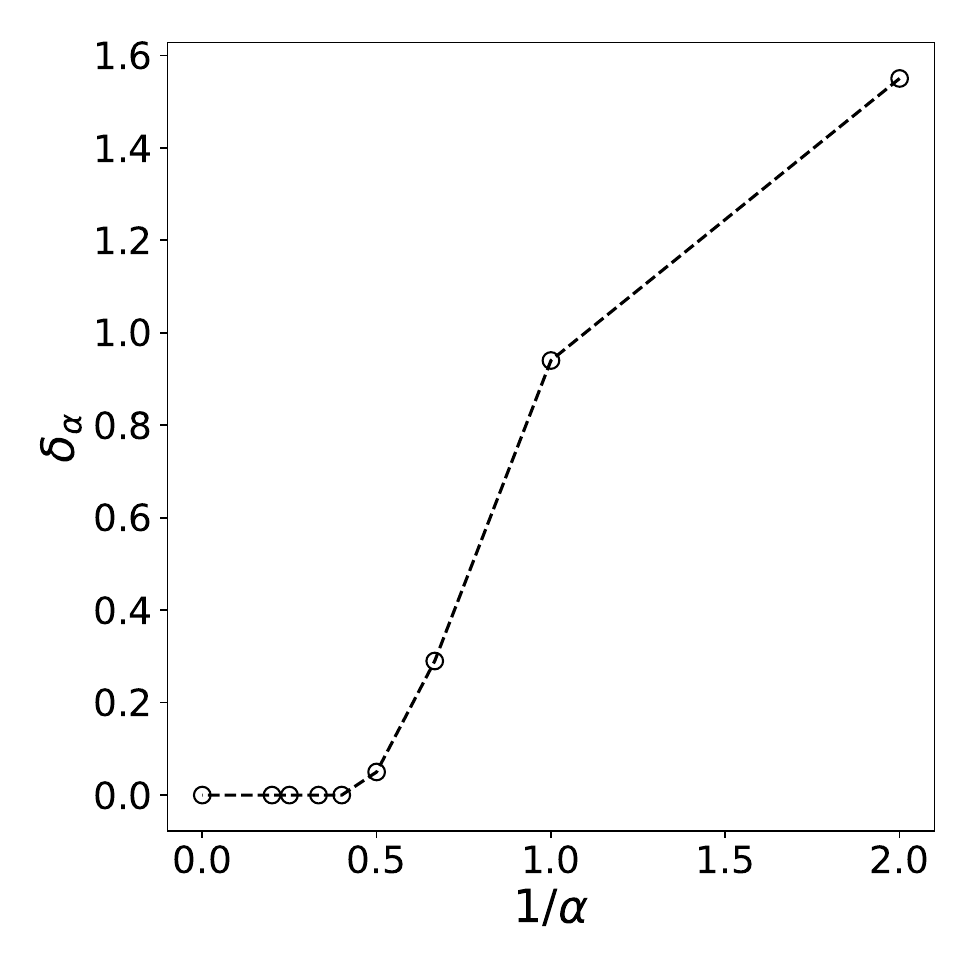}\\
 \includegraphics[width=5cm]{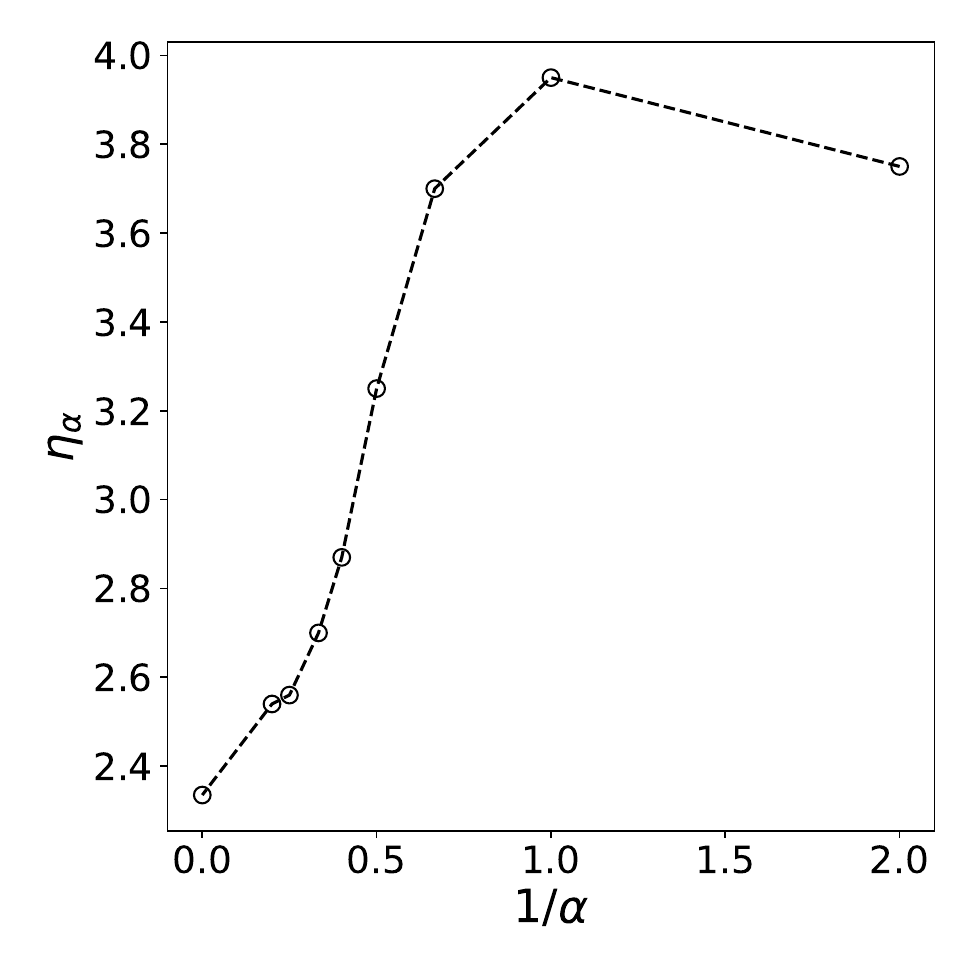}
 \includegraphics[width=5cm]{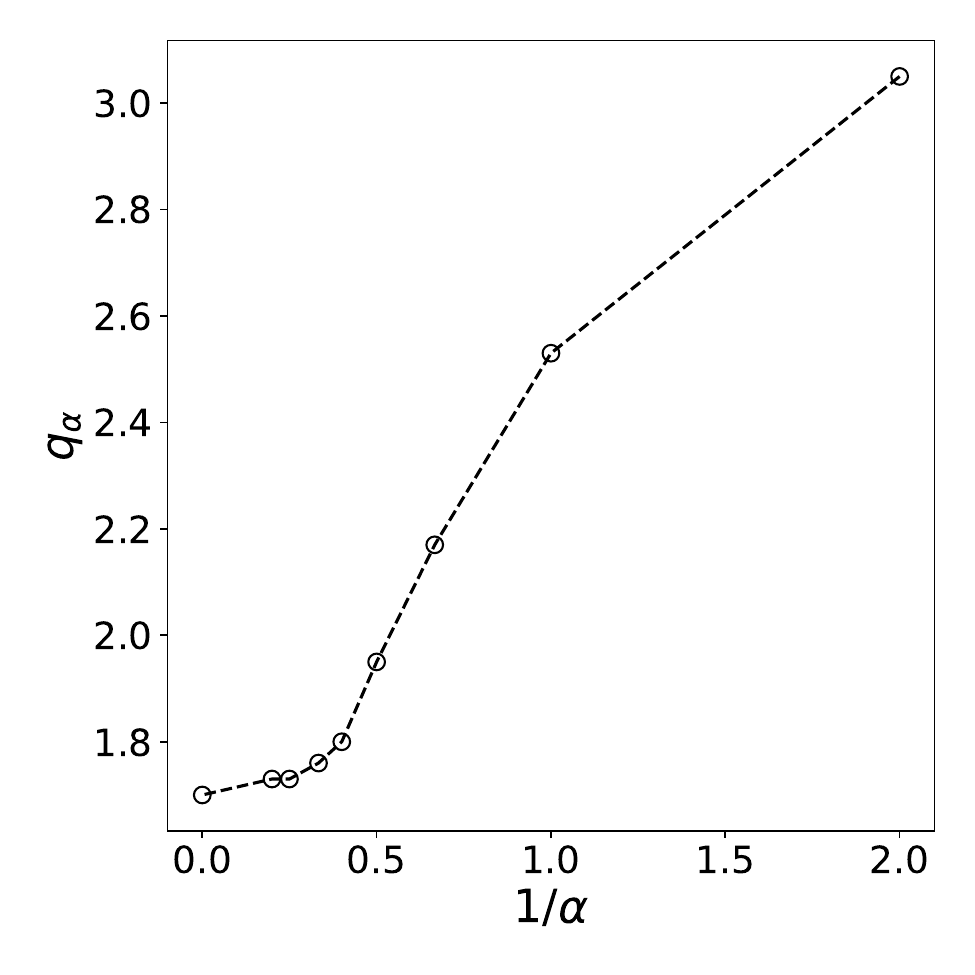} 
	\caption{Same data as in Fig.~\ref{alpha} versus $1/\alpha$.}
	\label{invalpha}
\end{figure*}

\begin{figure*}
	\centering
 \includegraphics[width=8.0cm]{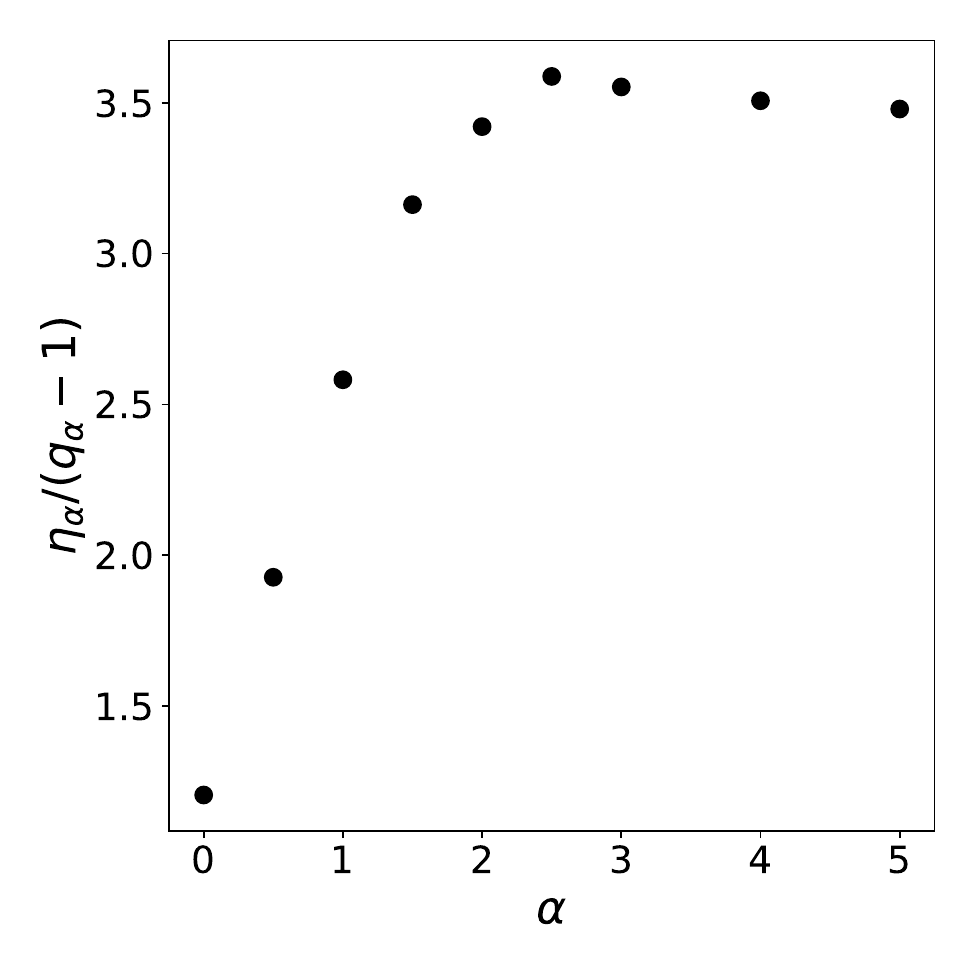}
	\includegraphics[width=8.0cm]{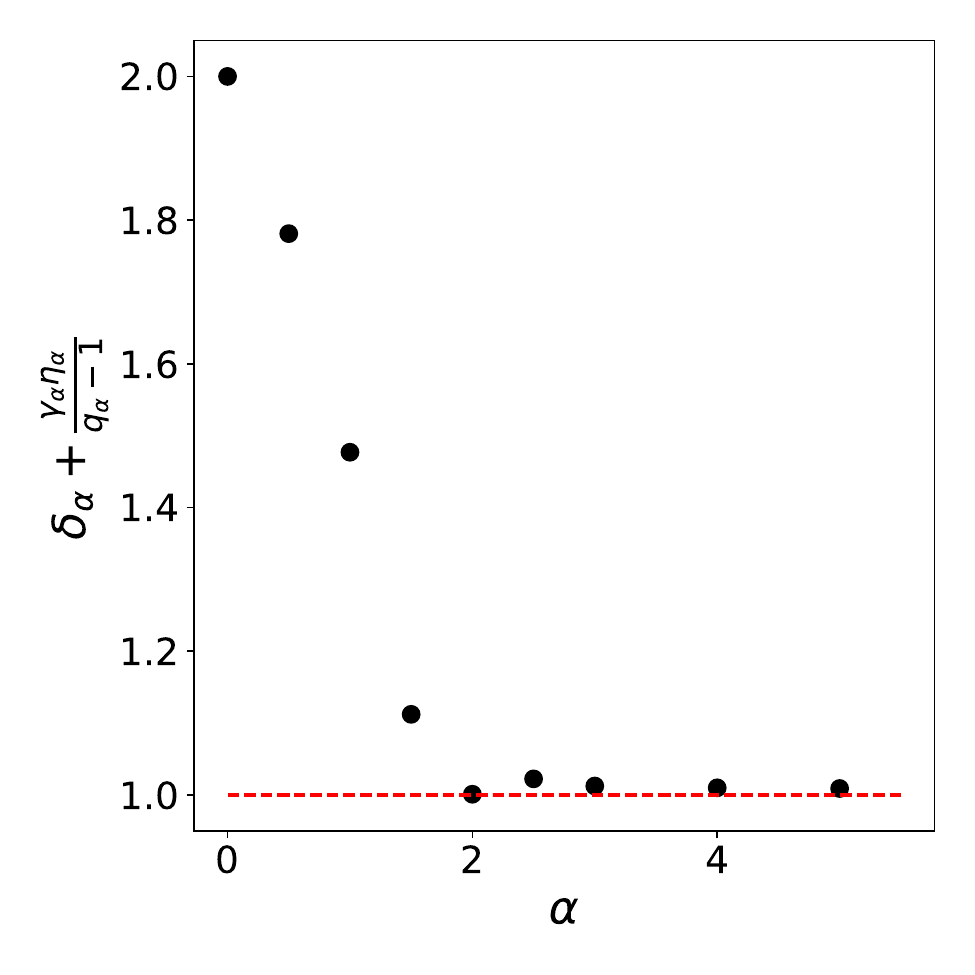}
   \caption{The $\alpha$-dependence of  the exponent of the large-temperature asymptotic behavior of $\sigma_{\alpha} \equiv \kappa/L \propto T^{-\eta_{\alpha}/(q_{\alpha}-1)}$ ({\it left}), and the exponent of the large-$L$ asymptotic behavior, $\sigma_{\alpha}\propto L^{-[\delta_\alpha+\gamma_\alpha \eta_\alpha/(q_\alpha -1)]}$ (\textit{right}). The requirement $\delta_\alpha+\gamma_\alpha \eta_\alpha/(q_\alpha -1)=1$ (red line) for the Fourier's law validity is numerically satisfied for $\alpha\ge 2$ and violated for $0 \le \alpha <2$.}
	\label{alphaslope}
  \end{figure*}

Let us discuss the values of the parameters of Eq.~\eqref{sigma}. As we see in Fig.~\ref{alpha}, the parameters $\gamma_{\alpha},\delta_{\alpha},q_{\alpha}$ are monotonic functions of $\alpha$. The parameter $\eta_{\alpha}$ exactly exhibits a peak at $\alpha_c$ and decreases after that. A curious fact is that $\delta_{\alpha}$ goes to zero as $\alpha>2$. This analysis, however, does not allow us to note the point at $\alpha\to\infty$. For this reason, let us consider the results in Fig.~\ref{invalpha}, which shows the parameters versus $1/\alpha$. $\eta_\alpha$ presents a peculiar behavior  at $\alpha_c=1$. In this particular analysis, however, $\gamma_{\alpha},\delta_{\alpha},q_{\alpha}$ exhibit atypical behavior around the neighbors of $\alpha_c$.

The thermal conductance scales like $L^{2-\alpha}$ ($\delta_{\alpha}=2-\alpha$ for $\alpha<\alpha_c$) in the range of \(0 \le \alpha < 1\). This can be justified by the derivation of the heat flux using
the energy of each rotor, which scales like energy per length. By using the property that $\mathcal{H}=\tilde{\mathcal{H}}/\tilde{L}$, where the renormalization is made by assuming a scale $t\to \sqrt{\tilde{L}}t$, a flux dependent on $L \tilde{L}$ is obtained. Without loss of generality, the following asymptotic expression can be used
\begin{equation}
    L\tilde{L}\sim \int_{1}^{\infty}dr\,\frac{r}{r^{\alpha}}=\frac{L^{2-\alpha}-1}{2-\alpha}\,,
    \label{newn}
\end{equation}
where it is assumed that $L\sim \int_{1}^{L}dx$, in the thermodynamic limit ($L-1\sim L$). Therefore, the  critical value, $\alpha_c^{*}$, related with Fourier's law validity threshold for this one-dimensional system is found to be $\alpha_c^*=2$ (Fig.~\ref{alphaslope}).  This value is physically and mathematically quite different from the critical value for the interactions, $\alpha_c$. From the physical perspective, this is related to the regime in which Fourier's law starts to be valid, while the previous critical value of $\alpha_c=1$ is related to the regime, up to that, where the model starts to be long-ranged or short-ranged. Below $\alpha_c=1$, the model is considered very long-range ($0\le \alpha<1$). The critical value of $\alpha_c^{*}=2$ can be explained in the following way: for $\alpha<\alpha_c^{*}$ the high temperature thermal conductivity becomes dependent on the lattice size. In the $\alpha>\alpha_c^{*}$ regime, the high temperature thermal conductivity becomes independent of the lattice size. Also about the exponents, still in the regimes of very long-range interactions  and high temperatures, the absolute value of the lattice size exponent can be obtained as $\delta_{\alpha}+\gamma_{\alpha}\frac{\eta_{\alpha}}{q_{\alpha}-1}=\frac{2-\alpha}{2}$. It can be obtained by the renormalized heat flux assuming straightforwardly a time scale $1/\sqrt{L\tilde{L}}\sim L^{-\frac{2-\alpha}{2}}$. It corroborates with the values of the exponents, for instance, for  $\alpha=0,0.5,1$, we obtain numerically the absolute values of the exponents as $\delta_{\alpha}+\gamma_{\alpha}\frac{\eta_{\alpha}}{q_{\alpha}-1}\sim 1, 0.78,0.48$ as compared to  $1,0.75,0.5$ (see Fig.~\ref{alpha}). These latter values correspond to the regime of very-long ranged interactions; they are not valid for the  intermediate region $1\le \alpha<2$.

We can see in Fig.~\ref{alphaslope} that Fourier's law is broken in the range of $0\le\alpha< 2$ while for $\alpha\ge 2$ the thermal conductivity no more is lattice size dependent. The exponent of the temperature increases from $\alpha=0$ to $\alpha=2$ and saturates for $\alpha>2$. However, the exponent of the lattice size decreases from $\alpha=0$ to $\alpha=2$ and saturates for $\alpha\le2$. It indicates that the requirement $\delta_{\alpha}+\gamma_{\alpha}\frac{\eta_{\alpha}}{q_{\alpha}-1}=1$ starts to be satisfied at this limit. For $\alpha>\alpha_c^{*}$, at the thermodynamic limit, the expression $(L^{2-\alpha}-1)/(2-\alpha)\sim 1/(\alpha-2)$, then, $\delta_{\alpha}=0$, as the scaling for the thermal conductivity at high-temperature regime, which yields a factor $\frac{1}{\sqrt{L\tilde{L}}}\sim \sqrt{\alpha-2}$. 

As previously mentioned, the low-temperature regime of the thermal conductivity, for $\alpha>2$ is proportional to the linear size $L$. It also can be explained with the aid of Eq.~\eqref{newn}. Let us notice that, at these values of $\alpha$, Eq.~\eqref{newn} becomes $L\tilde{L}\sim (\alpha-2)^{-1}$, thus yielding $\kappa_{\alpha>2}/L\propto (\alpha-2)$, therefore  $\kappa\propto L$. The very-long ranged regime at low temperature behaves as $\kappa_{0\le\alpha<1}\propto L^{-(1-\alpha)}$ which is the well-known asymptotic behavior of $1/\tilde{L}$. Although it is an interesting fact, the low-temperature regime is only completely understood by a quantum mechanical approach. However, it is worth highlighting that there are real systems with large spins which can therefore  be modeled quasi-classically, e.g., single-molecule magnets ~\cite{Hicke}. Notice that, in the low temperature regime, the system behaves like a set of harmonically coupled oscillators (see \cite{Defaveri2022}).

Let us emphasize that, the approach used here to obtain the scale of the thermal conductance at low and high temperatures and the scale for the thermal conductance can be extrapolated for the $d$-dimensional $\alpha-XY$ model in the regime of very long-range interactions. Here, we are assuming a heat flux along one direction. However, the energy flux leaving the hot reservoir initially behaves as if there was spatial isotropy, and only later on it focuses the cold reservoir. Consistently,  
the thermal conductance for low temperatures scales as $\int_1^{N^{1/d}} dr r^{d-1}r/r^{\alpha}\sim L^{d+1-\alpha}$, where $N=L^d$.  Similarly, the large-$L$ exponent of the thermal conductivity for higher temperatures can be predicted to scale as $d(1-\alpha/(d+1))$, based on the possible behavior of the mean field model ($\alpha=0$), which shares the same exponent for both regimes. Although the scale of the thermal conductivity coincides with the square root of $L\tilde{L}$, it leads to non-integer values of the scale of the thermal conductivity for this particular case. So, we assume that the $1/2$ in the denominator of this scale is because the denominator for a $d$-dimensional lattice is simply $d+1$. For instance, the thermal conductivities of the mean-field model, at low temperatures, behave as $L^{-2}, L^{-3}$ for two and three-dimensional systems, respectively. The high-temperature thermal conductivities of the same model behave as $L^{-2}, L^{-3}$, in agreement with the one-dimensional system, which has the same exponent at low and high temperatures. All exponents recover the limit $L^{0}$ when $\alpha>\alpha_c^{*}=d+1$ (see Fig. \ref{alphaslope} for $d=1$). Therefore, when Fourier's law starts to hold, it is expected that the  $d$-dimensional $\alpha-XY$ model possibly obeys these scales. The physical interpretation of the extra dimension in the scaling can be possibly related to circular waves. The $\alpha-XY$ model is a set of planar rotators that oscillates, but can not be interpreted as particles slowly carrying heat. The circular motion of the spins, makes the system propagate heat throughout the bulk, and, at very long-range regimes, it occurs in a large part of the system, in contrast with the system at $\alpha>1$, where the interaction becomes weak. For the short-range case ($\alpha\to\infty$), it ceases, staying only close to its neighbors.  

\section{Final remarks}
In summary, Fourier's law is a remarkable relation between the heat flux and the thermal gradient. Although it is well-verified in a wide class of models, including ferromagnetically coupled spin models, we can establish a limit for its validity. When $0\le \alpha <2$, we show numerically that this law is no longer obeyed, whereas it is obeyed  for $\alpha\ge 2$. Particularly for the mean field model, such a system is a perfect thermal insulator, which means that, in the thermodynamic limit, the thermal conductivity rapidly vanishes. However, for $\alpha\ge 2$ the lattice size exponent of the thermal conductivity is zero, indicating that the system is not dependent on the lattice size at high temperatures. The thermal conductance scale of the very long-ranged models can be obtained as a power law of the lattice size, whose exponent is $\delta_{\alpha}=2-\alpha$. The high-temperature exponent for the thermal conductivity allows us to write the relation $\frac{\delta_{\alpha}}{2}+\gamma_{\alpha}\frac{\eta_{\alpha}}{q_{\alpha}-1}=0$, which is also applicable for the very long-ranged cases. The relations obtained here can be extrapolated to higher dimensions. For instance, the validity of Fourier's law for two and three-dimensional generic-ranged systems is possibly $\alpha_c^{*}=3$ and $\alpha_c^{*}=4$ ($\alpha_c^{*}=d+1$ for a $d$-dimensional system), respectively. The extrapolation allows us to obtain the general $\delta_{\alpha}$ for a $d$-dimensional system in the very long-range regime, namely $\delta_{\alpha}=d+1-\alpha$, as well as the exponent of the thermal conductivity for the high-temperature regime ($\kappa_{\alpha(d)}\propto  L^{-d[1-\alpha/(d+1)]}$), which is consistent with the simplest case, $\alpha=0$. 
The scaled thermal conductance was well-fitted by a $q$-stretched exponential, a typical function of $q$-statistics, and it was proved useful in the context of magnon heat transport \cite{TsallisLima2023, LimaTsallis2023, LimaTsallisNobre2023}. Here and in the previous works referring to $n$-vector models, we were capable of obtaining closed expressions for the thermal conductance, and hence, the asymptotic limit of the thermal conductivity, then, verifying that classical inertial $n$-vector models present normal heat conduction. From a future perspective, we can numerically verify the regime of the validity of Fourier's law for the $d=2,3$ classical inertial $\alpha-XY$ models, as well as the possible agreement with our extrapolations.

\section{Acknowledgments}
We benefited from fruitful comments by E.P. Borges as well as partial financial support from CNPq and FAPERJ (Brazilian agencies). We also acknowledge LNCC for allowing us to use the Santos Dumont supercomputer. 
U.T. is a member of the Science Academy, Bilim Akademisi, Turkey.


\end{document}